*Advancing from phenomenological to predictive theory of ferroelectric oxide solution properties through consideration of domain walls*

*Atanu Samanta[1], Suhas Yadav[1], Or Shafir[1], Zongquan Gu[2], Cedric J.G. Meyers[2,3], Liyan Wu[2], Dongfang Chen[2], Shishir Pandya[4], Robert A. York[3], Lane W. Martin[4,5], Jonathan E. Spanier[2] and Ilya Grinberg[1]\**

[1]Department of Chemistry, Bar-Ilan University, Israel

[2]Department of Mechanical Engineering and Mechanics, Drexel University, Philadelphia PA, USA

[3]Department of Electrical and Computer Engineering, University of California at Santa Barbara, Santa Barbara, CA, USA

[4]Department of Materials Science and Engineering, University of California at Berkeley, Berkeley, CA, USA

[5]Materials Sciences Division, Lawrence Berkeley National Laboratory, Berkeley, CA, USA

*corresponding author, email:* ilya.grinberg@biu.ac.il

**Abstract:**

Prediction of properties from composition is a fundamental goal of materials science and can greatly accelerate development of functional materials.[1–8] It is particularly relevant for ferroelectric perovskite solid solutions where compositional variation is a primary tool for materials design. To advance beyond the commonly used Landau-Ginzburg-Devonshire and density functional theory methods that despite their power are not predictive, we elucidate the key interactions that govern ferroelectrics using 5-atom bulk unit cells and non-ground-state "defect-like" ferroelectric domain walls as a "simple as possible but not simpler" model systems. We also develop a theory relating properties at several different length scales that provides a unified framework for the prediction of ferroelectric, antiferroelectric and ferroelectric phase stabilities and the key transition temperature, coercive field and polarization properties from composition. The elucidated physically meaningful relationships enable rapid identification of promising piezoelectric and dielectric materials.

**Popular Summary:**

Prediction of properties from composition in terms of atomic properties is a fundamental goal of materials science and can greatly accelerate development of ferroelectric perovskite oxide materials that form the core of various sensor devices. Landau-Ginzburg-Devonshire (LGD) theory and density functional theory (DFT) calculations have been the main theoretical methods for understanding ferroelectric solid solution systems. While LGD is a powerful tool that compactly describes the system in terms of polarization, the compactness comes at the price of providing only a phenomenological description of the material and makes LGD fundamentally incapable of predicting the changes in the properties due to compositional changes. At the other extreme, DFT provides a precise description of the desired properties for essentially any composition. However, the complexity of the wavefunction makes its interpretation difficult so that DFT calculations can accurately evaluate the material properties but do not in and of themselves provide insight for materials design. Here, we construct a a unified theoretical framework for relating properties starting at the sub-Å atomic electronic-structure and ending with the target technologically relevant macroscale collective properties, using ferroelectric domain walls as a model system that is complex enough to provide information about the energy landscape beyond the ground state that is relevant to the properties of interest and yet simple enough to obtain physical understanding. The elucidated physically meaningful relationships enable rapid identification of promising piezoelectric and dielectric materials.

**Article Text:**

1. Introduction

Prediction of properties from composition is a fundamental goal of materials science and the ability to do so accurately could greatly accelerate the design of new, high-performance materials.[9] Ferroelectric perovskite oxide solid solutions are an excellent example of systems where a rich variety of phases and behaviors can be obtained by compositional variation. For example, compositional modification of the $PbTiO_3$ system (both with lead and/or titanium substitution) has been used to achieve high performance for a variety of applications[11,12], motivating the long-standing interest in these materials, including the recent high-throughput and machine-learning efforts[16,21,22]. Since simple materials have been largely explored, recent material-design efforts for ferroelectric perovskites have shifted to the exploration of the vast space of possible multicomponent solid solutions [10,11,20,12–19]

Landau-Ginzburg-Devonshire (LGD) theory and density functional theory (DFT) calculations have been the main theoretical methods used to describe and understand ferroelectric solid solution systems. While LGD is a powerful tool for understanding material behavior by compactly describing the system in terms of the order parameter (polarization $P$ for ferroelectric materials), the compactness comes at the price of providing only a phenomenological description of the material that makes LGD fundamentally incapable of predicting the changes in the properties due to the changes in composition. At the other extreme, DFT calculations can obtain the electronic wavefunction of the complex system and provide a precise description of the desired properties for essentially any composition. However, the complexity of the wavefunction leads to the difficulty in interpreting the DFT results and makes DFT calculations the *in silico* equivalent of experimental measurements that can accurately evaluate the necessary properties but do not in and of themselves provide insight or guidance for materials design. Therefore, a predictive theory for understanding the impact of compositional variation materials on ferroelectric properties and guiding the design of new materials must be based on atomic properties and interatomic interactions in ferroelectric solid solutions. The complexity of the possible interactions in such solutions means that the material properties are controlled by multiple interdependent parameters, making the relationship between composition and properties highly non-linear. Thus, obtaining analytical formulae for prediction of multicomponent material properties from composition appears to be an impossible task and a unified theory for predicting the relative stabilities of the ferroelectric, paraelectric and antiferroelectric phases as well as the crucial Curie temperature ($T_c$), polarization and coercive field ($E_c$) parameters of FE systems based on the properties of the constituent atoms has not been developed to date.

The lack of such predictive theoretical understanding has motivated the recent effort to use high-throughput computing and machine learning for accelerated materials exploration and design[16,21,22]. However, the exploration of multicomponent solutions and the creation of large data sets suitable for accurate machine learning require the use of large supercells and many different compositions that can only be carried out with powerful high-performance computing resources. Even when high-throughput computation is available, the size of the computationally obtained data sets is relatively small and full treatment of quaternary and quintary phase diagrams is difficult due to its extremely high computational cost (Fig. 1a). Meanwhile, experimental data sets tend to be imbalanced due to the strong focus on known high-performance compounds. Such lack of data hinders the use of deep learning for materials design. Furthermore, while machine-learning approaches can identify previously obscure correlations and connections in a system, the physical interpretation and meaning of those connections must still be filtered through the understanding of a materials scientist to provide fundamental insight. Thus, accurate analytical models are still an attractive choice for predicting properties from compositions.

Despite thousands of studies carried out over the last 70 years, the physical understanding of how composition of ferroelectric solid solutions gives rise to the properties of the material such as polarization, transition temperature and coercive field and the relative stabilities of the ferroelectric, antiferroelectric and paraelectric phases is still incomplete. Due to the complex nature of ferroelectric solid solutions, only several distinct relationships are available to guide research and design of these materials while a unified theory that starts at the level of the ion electronic structure and ends with the bulk collective behavior has not been developed. Furthermore, the development of such a theory has not even been attempted, perhaps indicating that the development of such a theory is considered to be impossible.

Even in the more narrow but still important field of ferroelectric domain walls, to answer the relatively simple question of what domain energies can be expected for different compositions or strain states requires the use of DFT calculations that rapidly become intractable for solid solutions and do not in and of themselves provide any understanding of the underlying physics that control these domain walls.

Landau-Ginzburg-Devonshire theory has been the main theoretical approach for understanding and describing ferroelectric systems in terms of the *P* order parameter. However, the LGD approach suffers from several drawbacks. First, as a purely phenomenological method, it is incapable of guiding the design of new solid solutions because the LGD parameters must be rederived for each composition. Second, for problems involving spatial variation in polarization, the gradient parameters $g_{ij}$ are used to characterize the energy cost of *P* variation. These parameters are usually assumed to be weakly varying with changes in the system such as strain, temperature and compositional variation and are approximated by constant values. To the best of our knowledge, there is no model of interatomic interactions available for estimating g and for evaluating whether or not $g_{ij}$ can be expected to be constant for example with temperature or strain variation. Thus, it is unclear whether this assumption is justified. Finally, the LGD theory is inherently incapable of predicting the transition temperature of ferroelectrics because the transition temperature is a necessary parameter for the theory and is incapable of predicting nucleation barriers for switching of the material because the nucleation energy can be made arbitrarily small by choosing an arbitrarily small grid and nucleus step size.

Here, to address these issues we develop a unified theoretical framework for understanding the relationships between the composition and properties of ferroelectric materials including solid solutions by constructing analytical relationships between material characteristics at different scales, starting with atomic identity (sub Å-scale) and ending with the target technologically relevant (macroscale) properties (Fig. 1b). We study the interactions in ferroelectric 5-atom bulk unit cells and domain walls to reveal the key interactions and provide a unified description for the key properties of these materials. We show that

1) Standard LGD theory fails in the simple task of providing a physical description for the changes in DW energies with strain

2) A-B repulsion is the interaction that predominantly mediates the coupling of polarization in neighboring unit cells and governs the energy of 180º domain walls. Thus, the energy contribution governed by the gradient term is in fact present even in a single-domain state and single-domain 5-atom calculations can be used to analytically predict domain wall energies that require much larger (at least 40 atom) DFT calculations. The elucidation of the physical origin of the gradient term allow accurate estimation of $g_{ij}$ and the limits of their accuracy.

3) The consideration of the energetics and interatomic interactions for the 180º domain wall allows the identification of a simple structural criterion for the relative stability of the antiferroelectric versus ferroelectric phase in perovskite oxides.

4) Surprisingly, the 90º and 70º domain walls in ferroelectric materials are only weakly affected by the short-range interatomic interactions and are governed solely by the Coulomb interactions arising from polarization inhomogeneity.

4) The transition temperatures in ferroelectrics scale with the energy of the 71° domain wall, explaining the deviations from the previously suggested $T_c \sim P^2$ empirical relationship and enabling description of the Tc of different classes of FE solutions by a single expression (instead of using different proportionality constants for $T_c \sim P^2$ for different classes of solutions).

5) The intrinsic coercive field of the material (i.e. in the absence of grain boundaries and vacancies) can be estimated easily from the $T_c$ value.

6) $T_c$, $P$ and $E_c$ can be estimated from composition using a multi-level *analytical* model based on the elucidated physical relationships between the interatomic interactions, $P$, $T_c$ and $E_c$.

7) In addition to ionic size and displacement, valence plays an important role in determining the polarization and $T_c$ of ferroelectric solid solutions.

8) The ionic displacement of each B-site cation is quantitatively controlled by a combination of the atom electronegativity and ionic size

9) The understanding of the interatomic interactions in ferroelectrics that give rise to domain wall energies enables the prediction of Peierls potential flattening that should lead to easy domain oscillation in (Ba,Sr)TiO$_3$ films with larger Sr content. This prediction is confirmed experimentally.

10) B-cation valence likely also plays an important role in promoting the piezoelectric response of FE solutions, providing guidance for the design of new piezoelectric materials.

Taken together these results provide a unified theoretical framework for understanding and designing ferroelectric perovskite oxides.

*II. Relationships between properties at different length scales*

Polarization-electric field ($P$–$E$) hysteresis loops serve as the fingerprint of ferroelectric materials and are characterized by two basic collective macroscale properties: the remanent polarization $P_s$ and coercive field $E_c$. The transition temperature $T_c$ is another basic macroscopic collective property that sets the operational temperature range of ferroelectric devices[11]. Additionally, for piezoelectric applications, the $d_{33}$ coefficient is the most widely used measure of performance and has long been known to be inversely correlated with $T_c$, leading to the well-known problem of the trade-off between $d_{33}$ and $T_c$ and $E_c$ (Fig. 1c).

We set $E_c$ and $T_c$ as our target properties of interest (because unlike $P$, $E_c$ and $T_c$ cannot be obtained by first-principles calculations) and use several levels of intermediate properties to related them to solid-solution composition (Fig. 1b). Since the valence, size, and electronegativity of metal atoms are of fundamental importance for cation-oxygen bonding, we use these parameters as the first electronic-structure-based and sub-Å level characteristics. Microscopically, the most basic characteristics of a ferroelectric 5-atom unit-cell structure are the local ionic displacements ($D_i$) that give rise to spontaneous $P$ because the coupling between the displacements leads to the preference of the material for an ordered-polar structure at low temperature. Therefore, $D_i$ form the second, ion-based and Å-scale level of properties. Since $P$ is a collective property of the material, we use it as the third-level property. Since the local-ionic displacements in the paraelectric phase are disordered, the $T_c$ is likely to be related to the energy cost of the local misalignment of the dipoles formed by the cation off-center displacements. Since such misalignment is found at ferroelectric-domain walls, the values of the domain-wall energies $\sigma_\theta$ (where $\theta$ is the angle of the change in the $P$ vector at the domain wall) characterize the energy landscape relevant to the higher-energy states accessed in the high-temperature paraelectric phases and will provide information about $T_c$. Additionally, it is known that $E_c$ is related to $\sigma_\theta$. Therefore, we use $\sigma_\theta$ as the fourth-level, nm-scale properties that are directly related to the target $E_c$ and $T_c$ properties. Since the relationship between $E_c$ and $\sigma_\theta$ is known from previous work ($E_c \sim \sigma_\theta^2/P$),[25–29] we focus on the elucidation of the relationship between $\sigma_\theta$ and the lower-level $P$ and $D_i$.

### III. *Relating collective properties to structure through domain-wall energetics*

To understand the atomistic origins of the ferroelectric domain-wall energy, we study the energies of tetragonal 180° domain walls that are the simplest domain walls in perovskite ferroelectrics for BaTiO$_3$ (BTO), KNbO$_3$ (KNO), and PbTiO$_3$ (PTO) in a variety of strain states (Fig. 2). We find that standard Landau-Ginzburg-Devonshire (LGD) theory based on $P$ as the order parameter does not provide a physically meaningful picture for the strain variation effect on domain walls even for these simple systems and obtains the gradient energy parameter changes of approximately 800%, 600% and 40% for BTO, KNO and PTO, respectively (see SI, S1). Therefore, to understand the interactions in the material that control $\sigma_\theta$, we go one level deeper and consider the effects of the displacements of the individual cations $D_i$.

Here, it is important to distinguish between the different possible definitions of cation displacements. The displacements of the *A*- and *B*-site cations can be measured with reference to the center of mass of their O$_{12}$ and O$_6$ oxygen cages; these displacements change the *A*-O and *B*-O bond lengths and bond energies and generate polarization. We denote these displacements by $d_A$ and $d_B$, respectively. Since $d_B$ is defined with respect to the center of the O$_6$ cages, a shear of the O$_6$ octahedra *even without any shift of the A- and B-cations from the high-symmetry positions* will generate non-zero $d_B$ and local dipole moments while preserving the symmetric distances between the *A*- and *B*-cations (Fig. 2a). Alternatively, it is possible to define the *A*- and *B*-cation displacements with reference to the *A*- and *B*-site positions in the high-symmetry structure. We denote these displacements as $D_A$ and $D_B$. These displacements may not change the *A*-O and *B*-O distances in the case where the oxygen cages displace together with the cations, but will change the *A*-*B* distances and the *A*-*B* repulsive interactions if $D_A$ and $D_B$ are different (Fig. 2b). Thus, $d_A$ and $d_B$ should be used to describe the energetics of the *A*-O and *B*-O interactions, and $D_A$ and $D_B$ should be used to describe the energetics of the *A*-*B* interactions.

We first calculated $\sigma_{180}$ only allowing the cations to displace while freezing the oxygen ions at their high-symmetry positions (O-fixed) and analyzed the energetics of the system in terms of *A*- and *B*-site displacements using

$$\Delta E = \sum [-0.5 a_1 d_B^2 + 0.25 a_{11} d_B^4 + 0.5 b_1 d_A^2 + g(D_A - D_B)^2] \qquad \text{Eq. (1)}$$

where $\Delta E$ is the energy difference between the ferroelectric and paraelectric states. Since the driving force for ferroelectric distortion is controlled by changes in the bond lengths in the local *A*-$O_{12}$ and *B*-$O_6$ environments[30], the local energy due to the distortion that is represented by the first two terms is controlled by $d_A$ and $d_B$, and is parameterized by the material- and strain-state-specific constants $a_1$, $a_{11}$, and $b_1$. The last term in Eq. (1) represents the energy cost due to the repulsive interactions between the *A*- and *B*-site cations. This energy cost depends on the difference between $D_A$ and $D_B$ as parameterized by the $g$ constant and its form in Eq. (1) is derived by the Taylor expansion of the exponential repulsive interaction between the *A*- and *B*-site cations (see SI, S2-4). Thus, the first two terms of Eq. 1 give the local energy $U_{loc}$ while the last term represents the gradient-energy term $U_g$. In contrast to the LGD approach where a single-domain configuration does not contain any information regarding $U_g$, the more detailed atomistic approach contains $U_g$ information *even at the level of single-domain 5-atom calculations*. This is a crucial advantage of the atomistic approach. We fit $a_1$, $a_{11}$, $b_1$, and $g$ to the results of 5-atom DFT calculations (see SI, S3) and using the fit values for the O-fixed 180° domain-wall (DW) case, we obtained good agreement with the DFT O-fixed $\sigma_{180}$ values (Figs. 2c and d).

Consideration of the dipole-dipole interactions at the 180° domain wall shows that an attractive interaction will be present between the dipoles formed by the titanium (100) displacements on the two opposite sides of the domain wall. Thus, in the absence of *A*-site (*B*-site) displacement that couples to the *B*-site (*A*-site) displacements in the neighboring cells, the anti-parallel displacement pattern will be preferred as is in fact found by our DFT calculations (Fig. 2e and see SI, S6). This interaction decreases $\sigma_{180}$ and likely accounts for the overestimation of $\sigma_{180}$ by our model.

Comparison of the O-fixed $\sigma_{180}$ and fully relaxed $\sigma_{180}$ values show that oxygen relaxation is quite important for domain-wall energies, with much higher domain-wall energies (by a factor of 1.5-4) obtained for the O-fixed case. Examination of the relaxed DFT domain-wall structure (Fig. 2a) shows that the $O_6$ octahedra at the domain wall shear, substantially decreasing the displacement of the titanium cations at the domain wall relative to the *A*-site cation. We note that the local energy of the *B*-site displacement is controlled by $d_{Ti}$ that can be created by the $O_6$ shear mode. By contrast, the gradient-energy term in Eq. 1 is controlled by $D_{Ti}$ and is not affected by the $O_6$ shear mode. Thus, the activation of the $O_6$ shear simultaneously keeps the local energy $U_{loc}$ low because $d_{Ti}$ is approximately equal to the single-domain $d_{Ti}$ and keeps the gradient $U_g$ low due to the small $D_{Ti}$ and small change in the *A-B* cation distance. The energy cost of $O_6$ shear is low and thus, avoidance of the trade off between $U_g$ and $U_{loc}$ dramatically lowers the domain-wall energy relative to O-fixed $\sigma$.

Assuming that the energy cost of the shear mode is zero and that $D_{Ti}$ is equal to one half of the single-domain value $D_A$ value (see SI, S5), we evaluated $\sigma$ based on the $a_1$, $a_{11}$, $b_1$, and $g$ constants obtained from 5-atom calculations to obtain the lower-bound estimate of $\sigma_{180}$. Good agreement [Figs. 2c and d] is obtained between the model and DFT values with the underestimation accounted for by our neglect of the $O_6$ shear mode cost (which increases with higher strain).

Having elucidated the energetics of $\sigma_{180}$ in terms of *A*- and *B*-site displacements, we consider the tetragonal 90° and rhombohedral 71° domain walls that control the switching and $E_c$ in bulk ferroelectrics (see SI, S7). In contrast to the 180° domain walls, these domain walls exhibit dipole-dipole interactions that are less favorable for the domain wall then for the single domain, giving rise to a contribution to $\sigma_{90}$ and $\sigma_{71}$ that is proportional to $P^2$. Interestingly, the *A-B* cation repulsion energy for the O-centered 90° domain wall is slightly *lower* than that in the single-domain structure (see SI, S7). Thus, $\sigma_{90}$ is essentially due to dipole-dipole interactions and scales approximately as $P^2$. For the 71° rhombohedral-phase domain wall, the domain-wall energy $\sigma_{71}$ is dominated by the dipole-dipole interactions proportional to $P^2$ and the *A-B* cation repulsion contribution is slightly positive (see SI, S7). The dependence of $\sigma_{90}$ and $\sigma_{71}$ on the square of single-domain polarization means that these domain-wall energies will decrease with higher temperature due to the well-known decreases of material polarization at higher temperature.

The understanding of the domain-wall structure and energetics enables us to relate *P*, $T_c$, and $E_c$. For rhombohedral (tetragonal) ferroelectrics, the coercive field $E_c$ at the temperature *T* [$E_c(T)$] scales with $\sigma_{71}(T)^2/P(T)$ [$\sigma_{90}(T)^2/P(T)$], and since $\sigma_{71}$ ($\sigma_{90}$) is largely due to dipole-dipole interactions and scales with $P^2$, we obtain $E_c(T) \sim P^3(T)$. According to simple Landau theory $P(T) = (P_0^2/T_c)(T - T_c)^{1/2}$, where $P_0$ is the 0 K polarization. As discussed above, we expect $T_c$ to scale with the 0 K. As discussed above, we expect $T_c$ to scale with the 0 K $\sigma_{71}$ This relationship between $\sigma_{71}$ and $T_c$ is similar to the well-known linear relationship between the vacancy-formation energy and the melting temperature in metals. In both cases, the microscopic features of the material at 0 K ($\sigma_{71}$ and vacancy formation energy) measure the energy cost of the configuration that is dominant in the high-energy disordered phase and are therefore correlated with the temperature of the order-disorder phase transition. Since our analysis of domain-wall energies showed that $\sigma_{71}$ is dominated by the $P^2$ contribution, $T_c \sim \sigma_{71}(0\ K) \sim P_0^2$ and we obtain $E_c\ (300\ K) \sim (T_c - 300)^{1.5}$. This relationship shows a good fit to the observed $E_c$ and $T_c$ data (Fig. 3a). We note that this relationship allows an easy evaluation of the intrinsic $E_c$ of a material that is difficult to obtain either by direct measurements (due to the presence of imperfections such as grain boundaries, vacancies, and impurities) or by $P$ measurements due to the effects of the electrode-oxide interface on the measured $P$ values. By contrast, $T_c$ values are easy to measure and are only weakly affected by the imperfections in the material.

We now focus on the relationship between $P_0$ and $T_c$. As discussed above, $\sigma_{71}$ is dominated by the dipole misalignment cost ($\Delta E_{\text{dip}}$) that scales as $P^2$ with an addition of a small *positive* energy ($\Delta E_{\text{rep}}$) due to the $A$-$B$ cation repulsive interactions that changes slowly (essentially constant) with increasing $P_0^2$ for the $P_0$ values in the range of 0.35-0.9 $C^2/m^4$ as the PTO content of a solid solution is increased[§] (Fig. 3b and SI, S7). Therefore, for $P_0^2 > 0.3$ $C^2/m^4$, $T_c$ should scale as $c + \alpha P^2$, where $c$ and $\alpha$ are constants, in excellent agreement with the observed $Tc$ dependence on the calculated DFT $P_0$ (Fig. 3c). Since $\Delta E_{\text{rep}}$ is small, $T_c$ can be also fairly accurately estimated by a simple proportionality to $P^2$. Thus, the understanding of the energetics of domain walls enables a simple prediction of $\sigma$, $T_c$, and $E_c$ of a material based solely on its ground-state single-domain displacements and polarization, establishing the links between the lower-scale $D_i$ and $P$ properties, higher-level $\sigma$ properties and the bulk material $E_c$ and $T_c$ properties (Fig. 1a).

---

[§] It is clear that $\Delta E_{\text{rep}}$ will go to zero as $P$ goes to zero, but we do not have data to determine the functional dependence of $\Delta E_{\text{rep}}$ for $P^2 < 0.3\ C^2/m^4$.

### IV. *Predicting structure from composition*

Next, we develop an analytical model to predict the $D_i$ and $P$ intermediate-level properties from solid-solution composition and atomic properties. Here, we follow the approach of treating the solid solutions as perturbations of the PbTiO$_3$ structure and consider the factors governing the cation displacements of a ferroelectric solid solution, namely i) the bonding of the cation with its nearest-neighbor oxygen atoms, ii) the coupling of its displacement to the displacements of other cations, iii) the effects of the cation cage volume, and iv) the effect of O$_6$ tilting found in materials with low tolerance factor $t$ values. We first consider how the intrinsic preference of the cation for off-centering in the PbTiO$_3$ environment that arises from its cation-oxygen and is the most important factor governing $D_i$ can be predicted by the atomic properties. This preference is characterized by the $D_0$ parameters empirically estimated in previous work[30]. For transition metal cations with $d^0$ electronic configuration (*i.e.*, Sc, Ti, Zr, Hf, Nb, Ta, W, Mo) in O$_6$ octahedra, the second-order Jahn-Teller distortion is the origin of the lower energy of the distorted structure. Based on vibronic-coupling theory[31,32], it was shown that the square of the ionic displacement squared is proportional to the off-diagonal vibronic-coupling matrix element $V_{\text{TM–O}}$ and the crystal stiffness that can be reflected by ionic radius $R$ and to the transition metal-oxygen states eigenvalue difference reflected by the electronegativity of the cation. Therefore, we estimated $D_0$ using:

$$D_0^2 = c_0 + c_1 R^{-2} + c_2 \chi \qquad \text{Eq. (2)}$$

where $c_0$, $c_1$, and $c_2$ are constants. This equation provides a good fit for the $D_0$ values obtained from DFT calculations (Fig. 4a). For the *B*-site cations without valence *d* states substituted into PbTiO$_3$, the intrinsic displacement is zero and their displacements are due to coupling with the titanium cations. We find that $D_0$ of these cations is also a linear function of $\chi$ and $R$ (Fig. 4a inset).

Next, we construct a model for $D_i$ that only takes factors i) and ii) into account. Here we seek to use a functional form that contains as few adjustable parameters as possible. The $B$-O bonds are stiffer than the $A$-O bonds and, therefore, we expect that $D_A$ will be affected by both $A$- and $B$-site composition, while $D_B$ will be affected by the $B$-site composition only, as described by

$$D_{B,i} = D^0_{B,i} * D^0_{B,avg} / D^0_{Ti} \qquad \text{Eq. (3)}$$

$$D_{A,i} = (D^0_{A,i} + \kappa_1 * D^0_{B,avg}) / (D^0_{Pb} + \kappa_1 * D^0_{Ti}). \qquad \text{Eq. (4)}$$

We then use the $D_i$ values to obtain $P_0$ according to $P_0 = \sum_i Z_i^* D_i$ where $Z_i^*$ are the cation Born effective charges, and then use $T_c = \alpha P_0^2$ to fit the $\kappa_1$ and $\alpha$ parameter to the experimental $T_c$ data for the lead- and bismuth-based solid solutions in order to fit our model to a large number of compositions. The experimental $T_c$ values for lead-only and lead- and bismuth-based solid solutions versus the estimated $P_0$ obtained using Eqs. 2-3 with the fitted $\alpha$ and $\kappa_1$ (Fig. 3b). We observe that solid solutions of PbTiO$_3$ with the homovalent Bi$Me^{3+}$O$_3$ ($Me$ = Ga, Sc, In) and PbZrO$_3$ substituents exhibit consistently higher $T_c$ values that lie away from the trend formed by the solid solutions of PbTiO$_3$ with the heterovalent Bi$Me'Me''$O$_3$ and Pb$Me'Me''$O$_3$ substituents. This is due to the underestimation of the $P$ values by our model for the Bi$Me^{3+}$O$_3$-PbTiO$_3$ and PbZrO$_3$-PbTiO$_3$ as can be seen for the PbZr$_{0.5}$Ti$_{0.5}$O$_3$ system where the $P_0$ predicted by the model is smaller than the 0.76 C/m$^2$ DFT value. This suggests that the valence of the substituent cations affects the $D_i$ values in PbTiO$_3$-derived solid solutions and must be considered.

We now consider the interplay of the effects iii) and iv) with the valence of the Bi$Me'Me''$O$_3$ and Pb$Me'Me''$O$_3$ substituents. Simple crystal chemistry arguments and DFT calculations[33] (SI, S12) show that dispersal of the substituent units in the solid solution will be favoured by the substitution of homovalent Bi$Me^{3+}$O$_3$ and PbZrO$_3$ due to weaker cation-oxygen under-/over-bonding while clustering of the substituent units will be favoured for the heterovalent Bi$Me^{'}$Me$^{''}$O$_3$ and Pb $Me^{'}$Me$^{''}$O$_3$ substituents due to stronger cation-oxygen under-/over-bonding. DFT calculations for Bi$Me^{3+}$O$_3$-PbTiO$_3$ ($Me$ = Ga, Sc, In) also show that for isolated substituent cations, the displacements increase with increasing ionic size of $Me^{3+}$, consistent with the well-known effect of volume expansion on the $A$-site ionic displacement[34]. By contrast, for systems with the Bi$Me^{3+}$O$_3$ units in close proximity, increasing $R$ and lower $t$ tend to decreased the cation displacements due to induced O$_6$ rotations[15,34]. Since the volume increase effect is proportional to the Bi$Me^{3+}$O$_3$ content, the Bi$Me^{3+}$O$_3$-PbTiO$_3$ and Pb$Me^{4+}$O$_3$-PbTiO$_3$ solutions will exhibit a positive linear dependence of $D_i$ on the content of the substituted $Me$ ($x_{Me}$) and $R_{Me}$. Since O$_6$ rotation require the cooperative distortion of at least two O$_6$ octahedra, the negative O$_6$ tilting effect on $D_i$ will have a quadratic dependence on the content of the substituted $Me$ ($x_{Me}$) and $t$.

By contrast, since for the mixed-valence substitution, the Bi$Me^{'}$Me$^{''}$O$_3$ and Pb $Me^{'}$Me$^{''}$O$_3$ substituents units are favoured to be in close proximity, both the positive effect of the volume expansion and negative effect of O$_6$ rotation scale linearly with $x_{Me}$. Taking the coupled effects of ionic size and valence into account (see SI, S12), and fitting the parameters for the ionic-size effects to the experimental $T_c$ data for the Bi$Me^{3+}$O$_3$-PbTiO$_3$ and Bi$Me^{'}$Me$^{''}$O$_3$-PbTiO$_3$, $P_0 = \sum_i Z_i^* D_i$ and $T_c = c + \alpha P^2$, we find quantitative agreement between the model predictions for the $T_c$ value and the experimental data for the Bi$Me^{3+}$O$_3$-PbTiO$_3$ and Bi$Me^{'}$Me$^{''}$O$_3$-PbTiO$_3$ materials used as the training set *and* the lead-only materials used as the test set (mean absolute error (MAE) of 48 K) as well as between the model predicted and DFT $P_0$ values (Fig. 4c and SI, S12). A fairly good agreement with MAE of 69 K is obtained using the $T_c = \alpha P^2$ relationship. The good agreement between the experimental and predicted $T_c$ values and the predicted and DFT-calculated $P_0$ values indicates that our model for $D_i$ is accurate despite having few (4 or 5) adjustable parameters and making some perhaps simplistic assumptions. Thus, we have connected solid solution composition to the local structure $D_i$ properties through the atomic $\chi$ and $R$ characterstics and have achieved the prediction of $T_c$ (and therefore $E_c$) bulk macro-scale properties from the composition of the solid solution.

## V. Design of new materials with analytical model

We now demonstrate how our theoretical framework enables the rapid design of new piezoelectrics, antiferroelectrics, and dielectrics.

An examination of the literature for $d_{33}$ and $T_c$ (Fig. 1c, see SI, S20) shows that the BiScO$_3$-PbTiO$_3$ derived morphotropic phase boundary (MPB) materials examined over the course of the last two decades[23,24] exhibit improved combinations of $d_{33}$ and $T_c$ (Pareto frontier) compared to other MPB materials. Unfortunately, scandium-based systems are ill-suited for practical applications where bulk materials are required due to the prohibitive cost of scandium, necessitating the development of high-performance scandium-free MPB piezoelectrics. Thus, to demonstrate the application of our analysis method, we seek to identify new scandium-free materials with higher $T_c$ and $d_{33}$ that fall on the Pareto frontier exhibited by scandium-containing systems.

Our identification of bismuth substituent oxide clustering as the key difference between BiSc$^{3+}$O$_3$-PbTiO$_3$ and other solid solutions suggests that this is also the origin of the shifted Pareto frontier exhibited by the BiSc$^{3+}$O$_3$-PbTiO$_3$ MPB solid solutions. Recent work has shown that the piezoelectric response at the MPB is due to the electric-field- (or stress-) induced reversible phase transformation between the rhombohedral and tetragonal phases[35]. Thus, the higher $d_{33}$ for BiScO$_3$-derived materials compared to the Bi*Me'Me''*O$_3$-derived materials at the same $T_c$ (and therefore $P$) is due to the flatter barrier for the motion of the rhombohedral-to-tetragonal boundary during the reversible transformation. Our results suggest that the higher barrier for the reversible rhombohedral-to-tetragonal transformation of Bi*Me'Me''*O$_3$-derived MPB materials is due to the presence of a large amount of bismuth-rich local environments that act as pinning sites for the motion of the rhombohedral-to-tetragonal boundary. Thus, MPB bismuth-based materials with dispersed bismuth atoms due to the use of Bi*Me*O$_3$ end-member where the *Me* has the oxidation state of +3 are likely to exhibit both high $T_c$ and high $d_{33}$.

Using our model for $T_c$ and the relationship between $d_{33}$ and $T_c$ indicated by the dashed line in Fig. 1c, we predict a series of MPB scandium-free high-performance solutions with minimal cation-oxygen under-/over-bonding (Fig. 1c and SI, S20) on the more favorable Pareto frontier spanning the $T_c$ range from 300 to 550 °C, that is much broader than that that afforded by the current known scandium-based systems. Given the complex nature of these solid solutions, their identification by experimental or DFT means would require on the order of months to years. Using our model, this task took ~1 hour, demonstrating the power of the analytical understanding.

Furthermore, the relationships learned between the composition, structure and domain wall energies reveals the criterion for the relative stabilities of the ferroelectric and antiferroelectric phases. Our 180° DW model results show that if the atoms cannot displace to minimize the *A-B* cation repulsion, the antiferroelectric phase is more stable than the ferroelectric phase. Therefore, we find that for several known lead-based double perovskites and the PbZr$_{1-x}$Ti$_x$O$_3$ solid solution, the antiferroelectric phase is more stable when the $D_{B,avg}$ of the ferroelectric phase is smaller than 0.11 Å (Fig. 5a). Thus, the ferroelectric single-domain *B*-cation displacement can be used to determine the relative stability of the antiferroelectric and ferroelectric phases, enabling rapid screening of candidate energy storage materials close to the antiferroelectric-ferroelectric phase boundary.

Going beyond the original target of prediction of the phase stabilities and the $T_c$, $P$ and $E_c$ parameters, we consider the implications of our model for dielectric materials. For BTO, our model predicts that the energy of barium-centered 180° domain walls ($\sigma_{Ba}$) scales as $P^{1.9}$ while the energy of the titanium-centered 180° domain walls ($\sigma_{Ti}$) scales as $P^3$ (see Fig. 5b and SI, S11). This means that the application of tensile strain that decreases the $c/a$ lattice parameter ratio and $P$ will strongly increase $\sigma_{Ti}$ but will more weakly decrease $\sigma_{Ba}$. Thus, for a sufficiently small $P$ (due to a small $c$ lattice parameter, see SI, S11), the Peierls potential given by $\sigma_{Ti} - \sigma_{Ba}$ will become flat or close to flat, enabling rapid thermal oscillations of domain walls. Since our analytical model and DFT results show that $D_A$ and $D_B$ are controlled by the $c$ lattice parameter (see SI, S11), the flattening of the Peierls potential and thermal domain-wall fluctuations should also occur when the $c$, $P$, and $U_{loc}$ of BTO are decreased by mixing with SrTiO$_3$ to obtain Ba$_{1-x}$Sr$_x$TiO$_3$ (BST) solid solutions.

## *VI. Discussion and Conclusion*

We now discuss the broader implications of our work. First, we note that the strategy pursued here is similar to that of deep learning in that rather than using a the more usual direct approach of trying to model the target properties in terms of basic characteristics of the system, a network of several interconnected levels of features is constructed to predict the target properties. Second, to advance from phenomenological to predictive description of FE properties, we use ferroelectric domain walls as a model system that is complex enough to provide information about the energy landscape beyond the ground state that is relevant to the properties of interest and that points out the limitations of the LGD approach and yet simple enough to obtain physical understanding. Third, the use of such "simple as possible but not simpler" domain wall model systems also allows us to develop an analytical description using atomic-based order parameters and their interactions rather than using only the average, coarse-grain unit-cell polarization *P* as the key parameter. In machine learning language, these three approaches respectively capture non-linear relationships, provide more balanced information for the model and make the model more flexible. This suggests that problems for which analytical theory and understanding have been lacking but machine and deep learning techniques have been successful are in fact amenable to the use of analytical physically-based models by using multiple-levels of relationships and flexible fine-grain analytical forms.

Furthermore, the successful use of DW defect to characterize the interactions in the system important for collective behavior along with the previous similar use of the vacancy energy to predict melting points in metals suggests that the approach of studying the structure and energy of defect states at 0 K for understanding of collective properties at high temperature is general and applicable to wide variety of physical systems.

Our work demonstrates that accurate analytical prediction of properties of complex ferroelectric solid solutions including properties that have been incalculable using either phenomenological approaches or state-of-the-art atomistic methods can be achieved by creating a multi-level theory that can be obtained through elucidation of fundamental relationships between several levels of properties based on sufficiently complex model systems. We hope that our work will inspire the application of this approach to a broad range of multicomponent systems and target properties and functionalities.

**References**

1.  Von Rohr, F., Winiarski, M. J., Tao, J., Klimczuk, T. & Cava, R. J. Effect of electron count and chemical complexity in the Ta-Nb-Hf-Zr-Ti high-entropy alloy superconductor. *Proc. Natl. Acad. Sci. U. S. A.* (2016) doi:10.1073/pnas.1615926113.

2. Ji, H. *et al.* Computational Investigation and Experimental Realization of Disordered High-Capacity Li-Ion Cathodes Based on Ni Redox. *Chem. Mater.* (2019) doi:10.1021/acs.chemmater.8b05096.

3. Kaufmann, K. *et al.* Discovery of high-entropy ceramics via machine learning. *npj Comput. Mater.* (2020) doi:10.1038/s41524-020-0317-6.

4. Castelli, I. E. *et al.* Computational screening of perovskite metal oxides for optimal solar light capture. *Energy Environ. Sci.* (2012) doi:10.1039/c1ee02717d.

5. Lederer, Y., Toher, C., Vecchio, K. S. & Curtarolo, S. The search for high entropy alloys: A high-throughput ab-initio approach. *Acta Mater.* (2018) doi:10.1016/j.actamat.2018.07.042.

6. Rohrer, G. S. *et al.* Challenges in ceramic science: A report from the workshop on emerging research areas in ceramic science. *J. Am. Ceram. Soc.* (2012) doi:10.1111/jace.12033.

7. Kitchin, J. R. Machine learning in catalysis. *Nature Catalysis* (2018) doi:10.1038/s41929-018-0056-y.

8. Lee, S. *et al.* Systematic Band Gap Tuning of BaSnO3 via Chemical Substitutions: The Role of Clustering in Mixed-Valence Perovskites. *Chem. Mater.* (2017) doi:10.1021/acs.chemmater.7b03381.

9. Maddox, J. Crystals from first principles. *Nature* **335**, 201 (1988).

10. Rabe, K. M., Ahn, C. H. & Triscone, J.-M. *Physics of ferroelectrics: a modern perspective*. (Springer, 2007).

11. Zhang, S. & Li, F. High performance ferroelectric relaxor-PbTiO 3 single crystals: Status and perspective. *Journal of Applied Physics* (2012) doi:10.1063/1.3679521.

12. Bell, A. J. Ferroelectrics: The role of ceramic science and engineering. *J. Eur. Ceram. Soc.* (2008) doi:10.1016/j.jeurceramsoc.2007.12.014.

13. Rondinelli, J. M. & Spaldin, N. A. Structure and properties of functional oxide thin films: Insights from electronic-structure calculations. *Advanced Materials* (2011) doi:10.1002/adma.201101152.

14. Ahart, M. *et al.* Origin of morphotropic phase boundaries in ferroelectrics. *Nature* (2008) doi:10.1038/nature06459.

15. Benedek, N. A. & Fennie, C. J. Why are there so few perovskite ferroelectrics? *J. Phys. Chem. C* (2013) doi:10.1021/jp402046t.

16. Balachandran, P. V., Kowalski, B., Sehirlioglu, A. & Lookman, T. Experimental search for high-temperature ferroelectric perovskites guided by two-step machine learning. *Nat. Commun.* (2018) doi:10.1038/s41467-018-03821-9.

17. Liu, Q. *et al.* High-performance lead-free piezoelectrics with local structural heterogeneity. *Energy Environ. Sci.* (2018) doi:10.1039/c8ee02758g.

18. Tao, H. *et al.* Ultrahigh Performance in Lead-Free Piezoceramics Utilizing a Relaxor Slush Polar State with Multiphase Coexistence. *J. Am. Chem. Soc.* (2019)

doi:10.1021/jacs.9b07188.

19. Li, F. *et al.* Giant piezoelectricity of Sm-doped Pb(Mg1/3Nb2/3)O3-PbTiO3 single crystals. *Science (80-. ).* (2019) doi:10.1126/science.aaw2781.

20. Qiu, C. *et al.* Transparent ferroelectric crystals with ultrahigh piezoelectricity. *Nature* (2020) doi:10.1038/s41586-019-1891-y.

21. Xue, D. *et al.* Accelerated search for BaTiO3-based piezoelectrics with vertical morphotropic phase boundary using Bayesian learning. *Proceedings of the National Academy of Sciences of the United States of America* vol. 113 13301–13306 (2016).

22. Balachandran, P. V. *et al.* Predictions of new AB O3 perovskite compounds by combining machine learning and density functional theory. *Phys. Rev. Mater.* (2018) doi:10.1103/PhysRevMaterials.2.043802.

23. Eitel, R. E. *et al.* New high temperature morphotropic phase boundary piezoelectrics based on Bi(Me)O3-PbTiO3 ceramics. *Japanese J. Appl. Physics, Part 1 Regul. Pap. Short Notes Rev. Pap.* (2001) doi:10.1143/jjap.40.5999.

24. Zhang, S. *et al.* Growth and electric properties of MPB BiScO3-PbTiO3 thin films on La0.7Sr0.3MnO3-coated silicon substrates. *J. Am. Ceram. Soc.* (2010) doi:10.1111/j.1551-2916.2010.03630.x.

25. Grinberg, I. & Rappe, A. M. First principles calculations, crystal chemistry and properties of ferroelectric perovskites. *Phase Transitions* **80**, 351–368 (2007).

26. Wojdeł, J. C. & Íñiguez, J. Testing simple predictors for the temperature of a structural phase transition. *Phys. Rev. B* **90**, 14105 (2014).

27. Shin, Y.-H., Grinberg, I., Chen, I.-W. & Rappe, A. M. Nucleation and growth mechanism of ferroelectric domain-wall motion. *Nature* **449**, 881–884 (2007).

28. Merz, W. J. Domain formation and domain wall motions in ferroelectric BaTiO3 single crystals. *Phys. Rev.* (1954) doi:10.1103/PhysRev.95.690.

29. Abrahams, S. C., Kurtz, S. K. & Jamieson, P. B. Atomic displacement relationship to curie temperature and spontaneous polarization in displacive ferroelectrics. *Phys. Rev.* **172**, 551–553 (1968).

30. Cohen, R. E. Origin of ferroelectricity in perovskite oxides. *Nature* (1992) doi:10.1038/358136a0.

31. Bersuker, I. B. Vibronic (pseudo Jahn-Teller) theory of ferroelectricity: Novel aspects and applications. *Ferroelectrics* **536**, 1–59 (2018).

32. Polinger, V., Garcia-Fernandez, P. & Bersuker, I. B. Pseudo Jahn-Teller origin of ferroelectric instability in BaTiO3 type perovskites: The Green's function approach and beyond. *Phys. B Condens. Matter* **457**, 296–309 (2015).

33. Yadav, S. & Grinberg, I. First-principles study of the composition, cation arrangement, and local structure in high-performance Bi(Me3+)O3-PbTiO3 (Me3+ =Ga, Sc, In) ferroelectric solid solutions. *Unpublished*.

34. Bilc, D. I. & Singh, D. J. Frustration of tilts and A-site driven ferroelectricity in KNbO3-LiNbO3 alloys. *Phys. Rev. Lett.* **96**, 1–4 (2006).

35. Liu, H. *et al.* Role of Reversible Phase Transformation for Strong Piezoelectric Performance at the Morphotropic Phase Boundary. *Phys. Rev. Lett.* (2018) doi:10.1103/PhysRevLett.120.055501.

36. Gu, Z. *et al.* Resonant domain-wall-enhanced tunable microwave ferroelectrics. *Nature* **560**, 622 (2018).

**Author Contributions**

A.S. performed DFT calculations and analytical modeling of DWs and AFE. S.Y developed the structure-property correlations and the analytical model for predicting $P$ and $T_c$ from composition. O.S. revealed the correlation between atomic properties and cation displacements. S.P., Z.G., and L.W. carried out film fabrication and XRD measurements. C.J.G.M. performed microwave device fabrication and measurements. D.C. and L.W. carried out PFM measurements. Z.G. and J.E.S. carried out LGD modeling. R.A.Y., L.W.M. and J.E.S supervised the experimental work. I.G. conceived the project and supervised the theory work.

**Acknowledgements**

We would like to acknowledge Peter K. Davies for helpful discussions about perovskite oxide chemistry.

**Supplementary Materials:**

*Computational Methods S9 and S14*

*Figures S1-S23*

*Tables S1-S18*

*References (1-134)*

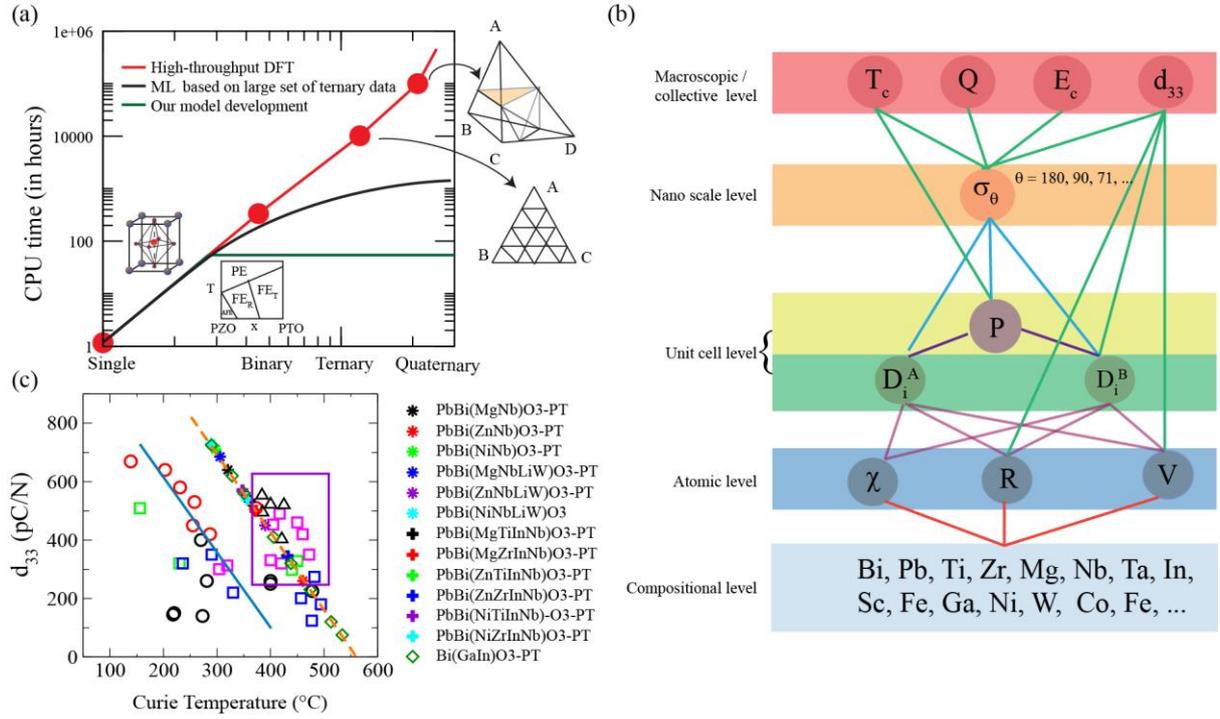

**Fig. 1. Unified multilevel theoretical framework for understanding and design of complex ferroelectrics.** (a) Schematic representation of lower limit for computational time required to predict material properties[§]. In practice, much higher (10-100 times) computational is likely to be incurred. The time required for experimental exploration of these solid solutions is likely to be greater by a factor of ~$10^3$-$10^4$. (b) Multilevel theory linking composition and electronic-structure sub-Å scale properties [electronegativity ($\chi$), ionic radius (R), and valence V] and macroscopic properties [such as polarization (P), domain-wall energy ($\sigma$), ferroelectric-to-paraelectric phase transition temperature ($T_c$), coercive field ($E_c$), piezoelectric constant ($d_{33}$), and quality factor (Q)]. (c) $d_{33}$ vs $T_c$ for Bi and Pb-based morphotropic phase boundary (MPB) systems. Black dashed line represents the trend for Sc-free solid solutions experimental data and the solid line shows the experimental trend for Sc-including solid solutions. Region with high values of $d_{33}$ and $T_c$ are shown by purple color box. The predicted Sc-free MPB solid solutions are shown by stars. The experimental data are taken from Supplementary Table S20.

[§] Total central processing unit (CPU) is calculated as $t_{CPU} = t_{scf} \times (n_1 \times n_2 \times n_3 \times 5)^3 \times {}^mC_p$, where, ${}^mC_p$ is a combination and $n_1$, $n_2$, $n_3$ are minimum dimensions of a suppercell required for studing a solid solution of pervoskite ABO$_3$ structure. Here, m=10 ( number elements), $t_{scf}$ = total CPU time required to finish a self consistent calculation of a five atom ABO3 pervoskite, and p = 2 for binary, p=3 for ternary and so on are considered for calculating $t_{CPU}$.

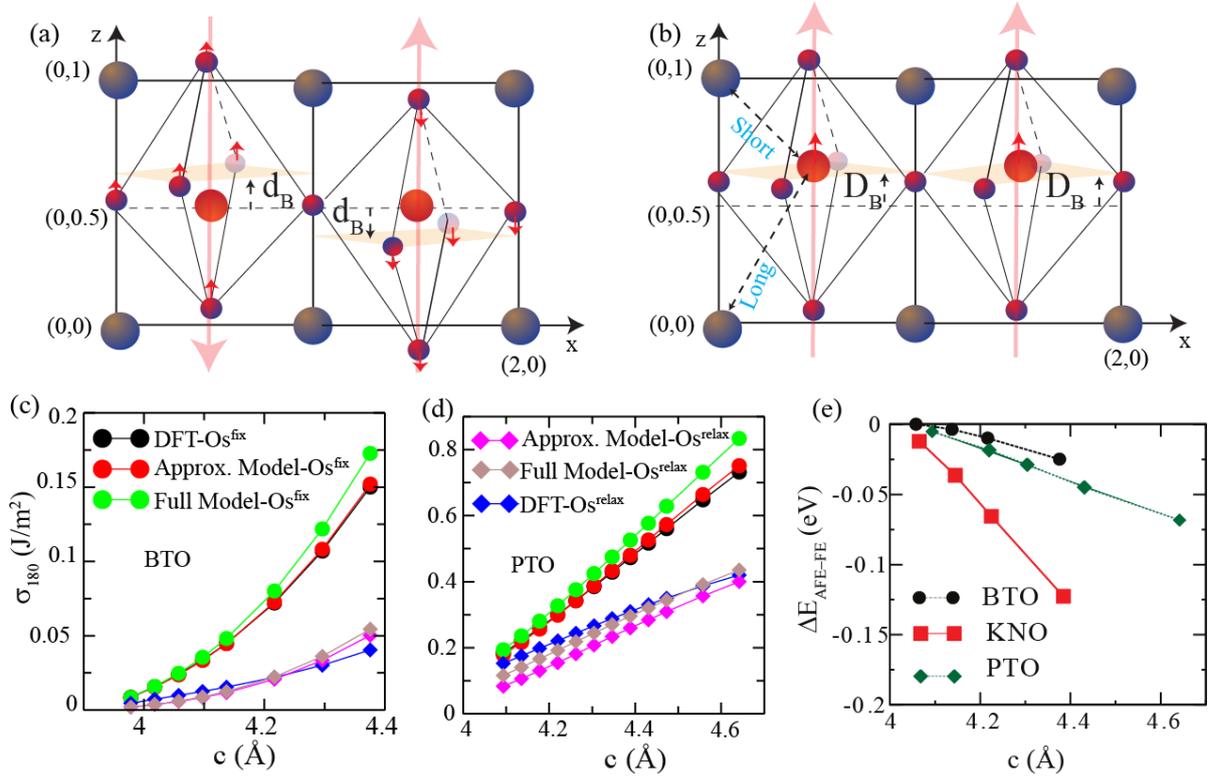

**Fig. 2. Interatomic interactions and domain wall energetics.** (a) and (b) Two-dimensional schematic representation of $ABO_3$ perovskite. (a) show tilting of $O_6$ oxygen cages without changing $A$-$B$ cation distance and displacement $d_B$ of $B$-cation atom with reference to the center of mass of their $O_6$ oxygen cages. (b) shows $B$-cation displacements $D_B$ with reference to the B-site positions in the high-symmetry structure. Similarly, we can define for $A$-cations, $d_A$ with reference to the center of mass of their $O_{12}$ oxygen cages and $D_A$ with reference to the A-site positions in the high-symmetry structure. (c) and (d) show the 180°DW energy as function of $c$ lattice parameter and share legends. (e) Energy difference between the antiferroelectric and ferroelectric state as a function of $c$ lattice parameter (the antiferroelectric and ferroelectric states are considered by keeping the O and $A$-site ions fixed at their symmetric positions).

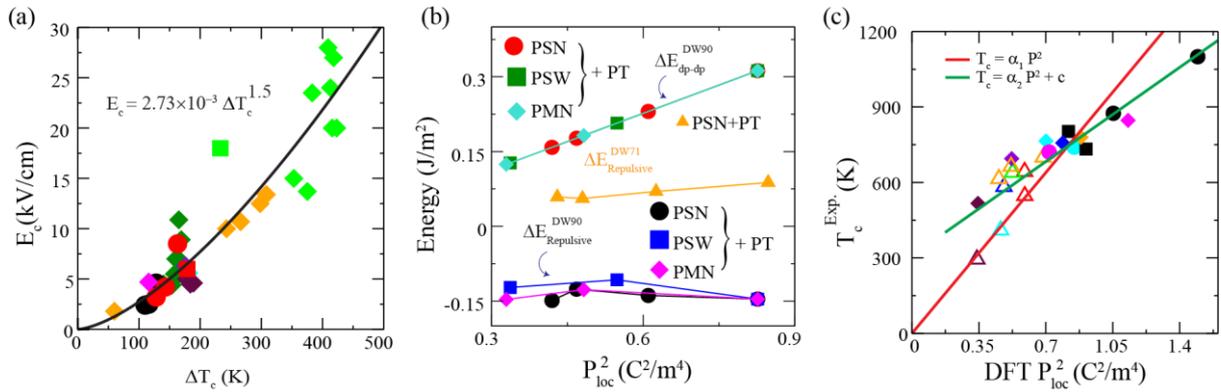

**Fig. 3. Relationships among the collective properties $P_{loc}$, $E_c$, and $T_c$.** (a) Reported experimental $E_c$ at 300 K versus $\Delta T_c$ (K) where $\Delta T_c = (T_c - 300)$ K. The correlation coefficient for the fit is 0.929. (b) Calculated dipole-dipole interaction and repulsive energy difference with respect to single domain as function of $P^2_{loc}$. The dipole-dipole interaction and repulsive energy are calculated using PTO model parameters. (c) Reported experimental value of FE to PE phase transition temperature ($T_c$) as function of square of reported DFT local polarizations ($P^2_{loc}$). The linear fitting constants $\alpha_1$, $\alpha_2$, and $c$ are 913.52, 535.23, and 308.43, respectively. The correlation coefficient for both linear fits is 0.897. References for experimental $T_c$, $E_c$, and DFT local $P$ data are given in S21.

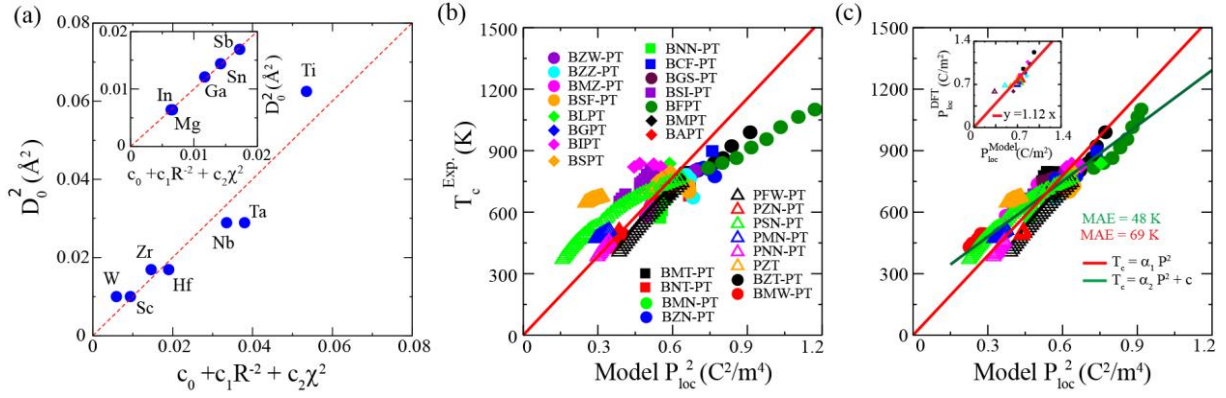

**Fig. 4. Relationship between atomic characteristics ($R, V, \chi$) and collective properties ($P_{loc}, T_c$).** (a) and its inset plot reported DFT $D^2_0$ value versus calculated $D^2_0$ using $D^2_0 = c_0 + c_1 R^{-1} + c_2 \chi^2$. Linear fit is shown by dotted red line. (b) Reported experimental $T_c$ versus model $P^2_{loc}$ without considering the effects of ionic size and valence. (c) $T_c$ versus $P^2_{loc}$ calculated considering effects of ionic size and valence parameters. The value of the optimized parameters $\kappa_1$, $\Lambda_1$, and $\Lambda_2$ is 27.47, 15.63, 45.49, respectively. The linear fitting constants $\alpha_1$, $\alpha_2$, and c are 1296.29, 902.91, and 209.19, respectively. The linear fit passing through origin and $T_c = \alpha_2 P^2_{loc} + c$ are shown by solid red and green lines. Top left inset shows DFT $P^2_{loc}$ versus model $P^2_{loc}$. The correlation coefficient for both linear fits is 0.897. References for experimental $T_c$, $E_c$, and DFT local $P$ data are given in S21.

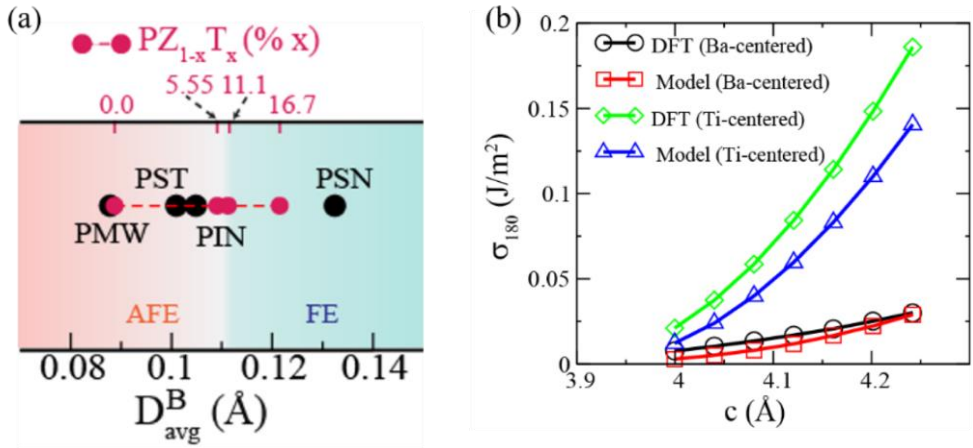

**Fig. 5. Prediction of material properties based on analytical model.** (a) Phase diagram of the antiferroelectric and ferroelectric phases as a function of average $B$-site displacement in a lead-based perovskite material. (b) $\sigma^{Ti}_{180}$ and $\sigma^{Ba}_{180}$ as function of $c$ lattice parameter of BTO (DFT and model calculations are considered using LDA functional and Eq.1 ).